\documentclass[aps,prb,twocolumn,superscriptaddress]{revtex4-2}

\usepackage{mhchem,graphicx,longtable,xfrac}
\usepackage[colorlinks=true,citecolor=blue,linkcolor=blue,breaklinks=true]{hyperref}
\usepackage{amssymb}
\usepackage{graphicx}
\usepackage{amsmath}

\usepackage{times}
\usepackage{color}
\usepackage{bm}

\begin{document}

\newcommand {\beq} {\begin{equation}}
\newcommand {\eeq} {\end{equation}}
\newcommand {\bqa} {\begin{eqnarray}}
\newcommand {\eqa} {\end{eqnarray}}
\newcommand {\ba} {\ensuremath{b^\dagger}}
\newcommand {\Ma} {\ensuremath{M^\dagger}}
\newcommand {\psia} {\ensuremath{\psi^\dagger}}
\newcommand {\psita} {\ensuremath{\tilde{\psi}^\dagger}}
\newcommand{\lp} {\ensuremath{{\lambda '}}}
\newcommand{\A} {\ensuremath{{\bf A}}}
\newcommand{\Q} {\ensuremath{{\bf Q}}}
\newcommand{\kk} {\ensuremath{{\bf k}}}
\newcommand{\qq} {\ensuremath{{\bf q}}}
\newcommand{\kp} {\ensuremath{{\bf k'}}}
\newcommand{\rr} {\ensuremath{{\bf r}}}
\newcommand{\rp} {\ensuremath{{\bf r'}}}
\newcommand {\ep} {\ensuremath{\epsilon}}
\newcommand{\nbr} {\ensuremath{\langle ij \rangle}}
\newcommand {\no} {\nonumber}
\newcommand{\up} {\ensuremath{\uparrow}}
\newcommand{\dn} {\ensuremath{\downarrow}}
\newcommand{\rcol} {\textcolor{red}}



\title{Magnetic properties and signatures of moment ordering in triangular lattice antiferromagnet KCeO$_2$}

\author{Mitchell M. Bordelon}
\affiliation{Materials Department, University of California, Santa Barbara, California 93106, USA}

\author{Xiaoling Wang}
\affiliation{Department of Physics and Center for Terahertz Science and Technology, University of California, Santa Barbara, California 93106, USA}

\author{Daniel M. Pajerowski}
\affiliation{Neutron Scattering Division, Oak Ridge National Laboratory, Oak Ridge, Tennessee 37831, USA}

\author{Arnab Banerjee}
\affiliation{4Department of Physics, Purdue University, West Lafayette, Indiana 47906, USA}

\author{Mark Sherwin}
\affiliation{Department of Physics and Center for Terahertz Science and Technology, University of California, Santa Barbara, California 93106, USA}

\author{Craig M. Brown}
\affiliation{Department of Chemical and Biomolecular Engineering, University of Delaware, Newark, Delaware 19716, USA}
\affiliation{Center for Neutron Research, National Institute of Standards and Technology, Gaithersburg, Maryland 20899, USA}

\author{M. S. Eldeeb}
\affiliation{Institute for Theoretical Solid State Physics, Leibniz IFW Dresden, Helmholtzstrasse 20, 01069 Dresden, Germany}

\author{T. Petersen}
\affiliation{Institute for Theoretical Solid State Physics, Leibniz IFW Dresden, Helmholtzstrasse 20, 01069 Dresden, Germany}

\author{L. Hozoi}
\affiliation{Institute for Theoretical Solid State Physics, Leibniz IFW Dresden, Helmholtzstrasse 20, 01069 Dresden, Germany}

\author{U. K. R\"{o}{\ss}ler}
\affiliation{Institute for Theoretical Solid State Physics, Leibniz IFW Dresden, Helmholtzstrasse 20, 01069 Dresden, Germany}

\author{Martin Mourigal}
\affiliation{School of Physics, Georgia Institute of Technology, Atlanta, Georgia 30332, USA}

\author{Stephen D. Wilson}
\email[]{stephendwilson@ucsb.edu}
\affiliation{Materials Department, University of California, Santa Barbara, California 93106, USA}

\date{\today}

\begin{abstract}
	The magnetic ground state and the crystalline electric field level scheme of the triangular lattice antiferromagnet KCeO$_2$ are investigated. Below $T_N =$ 300 mK, KCeO$_2$ develops signatures of magnetic order in specific heat measurements and low energy inelastic neutron scattering data. Trivalent Ce$^{3+}$ ions in the $D_{3d}$ local environment of this compound exhibit large splittings among the lowest three $4f^1$ Kramers doublets defining for the free ion the $J = 5/2$ sextet and a ground state doublet with dipole character, consistent with recent theoretical predictions in M. S. Eldeeb \textit{et al.} Phys. Rev. Materials 4, 124001 (2020). An unexplained, additional local mode appears, and potential origins of this anomalous mode are discussed.	
\end{abstract}


\maketitle

\section{Introduction}

Magnetically frustrated materials provide unique settings for realizing new electronic phases of matter. For instance, the triangular lattice antiferromagnet has been a prime platform for searching for highly entangled, quantum spin liquid (QSL) states and other unconventional ground states\cite{Lee, balents4, balents5, witczak6, zhou7, cava_broholm}. Antiferromagnetic nearest neighbor interactions across a triangular lattice network can preclude conventional magnetic order, depending on the strength of quantum fluctuations and exchange anisotropies. In the Heisenberg limit, the ground state of the nearest neighbor triangular lattice antiferromagnet is a three-sublattice order with moments rotated 120$^{\circ}$ relative to one another across a single triangle \cite{j1j2_1, j1j2_2, jolicoeur_bacci}. However, strong quantum fluctuations associated with small $S_{eff} = 1/2$ moments combined with strong anisotropies can instead favor a fluctuating QSL ground state such as the two-dimensional Dirac quantum spin liquid \cite{balents5, iaconis2018spin, bordelon2019field, PhysRevLett.123.207203} or ``resonating valence bond" states \cite{anderson1987resonating, anderson1973resonating, moessner_sondhi, NatComm8, kimchi2018valence, li2016muon}. 

While the triangular lattice is a relatively common lattice motif, few materials show signatures of hosting intrinsically quantum disordered magnetic ground states. Recently, several rare earth, Yb-based oxides were reported to form quantum disordered ground states, with prominent examples being YbMgGaO$_4$ \cite{NatComm8, kimchi2018valence, li2016muon, AdvQTech2, PRL122, zhang_mourigal, zhu_chernyshev, li_zhang2, li_zhang3, li_zhang4, paddison_mourigal, shen_zhao} and NaYbX$_2$ ($X$ = chalcogenide) \cite{bordelon2019field, PhysRevB.101.224427, baenitz_doert, ding_tsirlin, liu_zhangAMX2, ranjith_baenitz, ranjith_baenitz2, sarkar_klauss, sichelschmidt_doert, PhysRevMaterials.4.064410}. These and related materials do not form long-range magnetic order and instead display continuum spin excitations and thermodynamic properties consistent with QSL ground states. 

\begin{figure}
	\includegraphics[scale=.5]{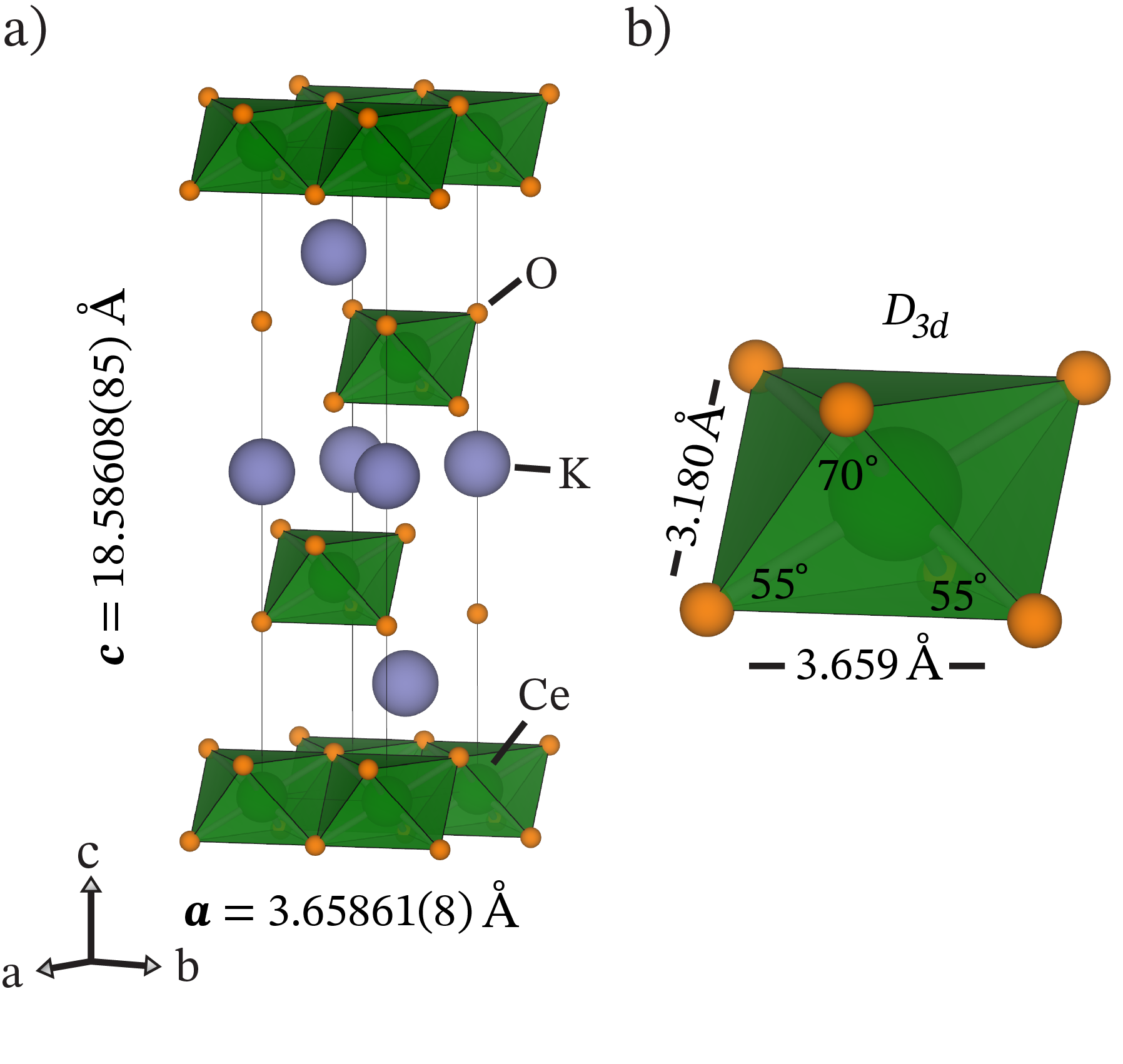}
	\caption{a) Experimentally determined lattice parameters and $R\bar{3}m$ crystal structure of KCeO$_2$ at 300 mK. Layers of K ions (purple) separate triangular sheets of CeO$_6$ octahedra. b) The CeO$_6$ octahedra are trigonally compressed along the crystallographic $c$-axis, resulting in a $D_{3d}$ symmetry local environment about the Ce ions.}
	\label{fig:structure_1}
\end{figure}


One of the appeals of using rare earth moments in strong ligand fields to explore the states possible within a triangular lattice is the ability to vary the character (\textit{e.g.} anisotropies) of the resulting $S_{eff} = 1/2$ moments via chemical substitution. Altering the rare earth ions can change moment size, $g$-factor anisotropy, exchange anisotropy, as well as the nature of the ground state multiplet. For example, swapping Tm$^{3+}$ for Yb$^{3+}$ in the YbMgGaO$_4$ structure results in multipolar, three-sublattice order \cite{shen_zhao_TMGO, li_gegenwart, cevallos_cava}. Another, recent rare earth QSL candidate, NaYbO$_2$, is a member of a much larger series of $ARX_2$ ($A =$ alkali, $R =$ rare earth, $X =$ chalcogenide) compounds, and a large number of these compounds with varying $R$-sites crystallize within the same $R\bar{3}m$ triangular lattice structure \cite{bordelon2019field, PhysRevB.101.224427, baenitz_doert, ding_tsirlin, liu_zhangAMX2, ranjith_baenitz, ranjith_baenitz2, sarkar_klauss, sichelschmidt_doert, PhysRevMaterials.4.064410, PhysRevB.103.014420, hashimoto2002structures, hashimoto2003magnetic, dong2008structure, cantwell2011crystal, xing-sefat2, fabry2014structure}. This invites exploration of the effect of tuning the character of the $R$-site magnetic ion in NaYbO$_2$ and exploring the response on its quantum disordered ground state.

Here we report a study of the single f-electron, Ce-based, analog of NaYbO$_2$ in the triangular lattice antiferromagnet KCeO$_2$. This compound possesses $S_{eff} = 1/2$ moments with a large $g$ factor anisotropy---one where the strong crystalline electric field surrounding trivalent Ce$^{3+}$ ions is on the same order of magnitude as the Ce ion's spin orbit coupling strength. We observe thermodynamic signatures of magnetic ordering, and, surprisingly, we also observe a well-separated extra crystalline electric field excitation that cannot be explained by traditional models of the Ce$^{3+}$ $J = 5/2$ ground state multiplet. We propose a crystalline electric field scheme consistent with a recently reported \textit{ab initio} model \cite{PhysRevMaterials.4.124001} and discuss the potential origins of the additional, anomalous $f$-electron mode.

\section{Methods}

\subsection{Sample preparation}

KCeO$_2$ was prepared by reducing CeO$_2$ with potassium metal adapted from a previous report \cite{clos1970deux}. A 1.1:1.0 molar ratio of K (99.95\% Alfa Aesar) and CeO$_2$ (99.99 \% Alfa Aesar) was sealed in 316 stainless steel tubing inside of an Ar filled glove box following similar syntheses of NaCeO$_2$ and NaTiO$_2$ \cite{PhysRevB.103.024430, wu2015natio}. The sealed stainless steel tubes were placed in an actively pumped vacuum furnace at 800 $^{\circ}$C for three days and opened in the glove box, revealing a highly air and moisture sensitive red-brown powder. All subsequent measurements were obtained with strict atmospheric control of the sample. Sample composition was verified via x-ray diffraction measurements under Kapton film on a Panalytical Empyrean powder diffractometer with Cu-K$\alpha$ radiation, and data were analyzed using the Rietveld method in the Fullprof software suite \cite{rodriguezfullprof} and GSAS/EXPGUI programs \cite{larson2004general, toby2001expgui}.  Powder samples were made in 5g batches. At most, only two batches were combined for powder neutron scattering experiments, where, prior to mixing, the samples were checked for quality and composition via x-ray diffraction and then thoroughly ground together. 

\subsection{Magnetic susceptibility, heat capacity, and electron paramagnetic resonance}

Magnetic properties of KCeO$_2$ were collected using a Quantum Design Magnetic Properties Measurement system (MPMS3) with a 7 T magnet and a Quantum Design Physical Properties Measurement System (PPMS) with a 14 T magnet and vibrating sample magnetometer. Zero field cooled (ZFC) and field cooled (FC) magnetic susceptibility of KCeO$_2$ from 2 K to 300 K were measured in the MPMS3 under an applied field of $\mu_0H = 0.5$ T. Isothermal magnetization data were collected at 2, 10, 100, and 300 K in the PPMS in fields up to $\mu_0H = 14$ T. Heat capacity was measured from 2 K to 300 K in the PPMS. At lower temperatures, heat capacity data were collected using a dilution refrigerator insert within the PPMS between 80 mK and 4 K in external fields of $\mu_0H = 0, 9,$and $14$ T. Electron paramagnetic resonance (EPR) spectra were collected at 4 K in an X-band EMXplus (Bruker) EPR spectrometer in the perpendicular operation mode, and data were modeled with the EasySpin package implemented in MATLAB \cite{stoll2006easyspin}.

\subsection{Neutron scattering} 
Neutron powder diffraction data were collected on the high-resolution powder diffractometer BT-1 at the National Institute of Standards and Technology (NIST) Center for Neutron Research (NCNR). The sample was loaded in a $^3$He cryostat and dilution refrigerator as pressed pellets in a copper can pressurized with He gas, and data were collected with incident neutrons of wavelength 2.0774 \AA \ using a Ge(311) monochromator. Structural analysis was performed with Rietveld refinement using the GSAS/EXPGUI program \cite{larson2004general, toby2001expgui}.  Low-energy inelastic neutron scattering (INS) data were collected using 10 g of KCeO$_2$ powder at the Cold Neutron Chopper Spectrometer (CNCS) instrument at the Spallation Neutron Source, Oak Ridge National Laboratory (ORNL).  The sample was loaded as pressed pellets into an 8 T cryomagnet with a dilution insert in a copper can with He exchange gas, and data were collected using incident neutrons of $E_i$ = 3.32 meV. The magnet and instrument background were approximated by subtracting empty copper can scans at 1.8 K. High-energy INS data were collected at the wide Angular-Range Chopper Spectrometer (ARCS) at the Spallation Neutron Source. 5 g of powder were loaded within an aluminum can closed-cycle refrigerator and data were collected at 5 K and 300 K. Two incident energies of $E_i$ = 300 meV (Fermi 1, Fermi frequency 600 Hz) and $E_i$ = 600 meV (Fermi 1, Fermi frequency 600 Hz) were used, and background contributions were removed by measuring an empty sample can at the same incident energies and temperatures.

\subsection{\textit{Ab initio} modeling} 
Modeling of the energy spectrum of defect states in KCeO$_2$ was performed using
complete active space self-consistent field
(CASSCF) quantum chemical computations \cite{Helgaker2000} including spin-orbit coupling (CASSCF+SOC). 
	An embedded-cluster material model was employed.
	With this approach, electronic-structure calculations are carried out just for a finite set of atoms
	(i.\,e., a cluster), while the remaining part of the extended crystalline surroundings is described
	as an effective electrostatic field.
	For the type of defect we focused on, a Ce$^{3+}$ ion within the K layer, a [(CeO$_6$)Ce$_{12}$K$_6$]
	cluster was considered.
	Effective core potentials and valence basis sets as optimized in Refs.\;\cite{ECPs_RE_1_dolg_89,
		BSs_RE_dolg_02} were used for the `central' Ce ion at the 3b crystallographic position, along with all-electron [4$s$3$p$2$d$] Douglas-Kroll basis sets for the adjacent ligands \cite{BSs_DK_deJong_01}.
	The positions of the latter were not reoptimized.
	To model the charge distribution in the immediate vicinity, we relied on large-core relativistic pseudopotentials including the 4$f$ electrons in the core as concerns the twelve Ce nearest neighbors \cite{ECPs_RE_2_dolg_89,BSs_RE_dolg_93} and on total-ion potentials as concerns the six adjacent K sites \cite{ECPs_fuentealba_82}.
	Charge neutrality was ensured through distributing a negative charge of $-2~\text{e}$ at metal sites far away from the defect Ce ion sitting at a 3b K site.
	An active space defined by the seven Ce 4$f$ orbitals was used in the CASSCF computation.
	The latter was performed for an average of the seven 4$f^1$ $S\!=\!1/2$ states.
	Spin-orbit coupling was subsequently accounted for according to the procedure described in \cite{SOC_molpro}.
	The quantum chemical package {\sc molpro} \cite{molpro12} was employed for these simulations.

\begin{figure}
	\includegraphics[scale=.55]{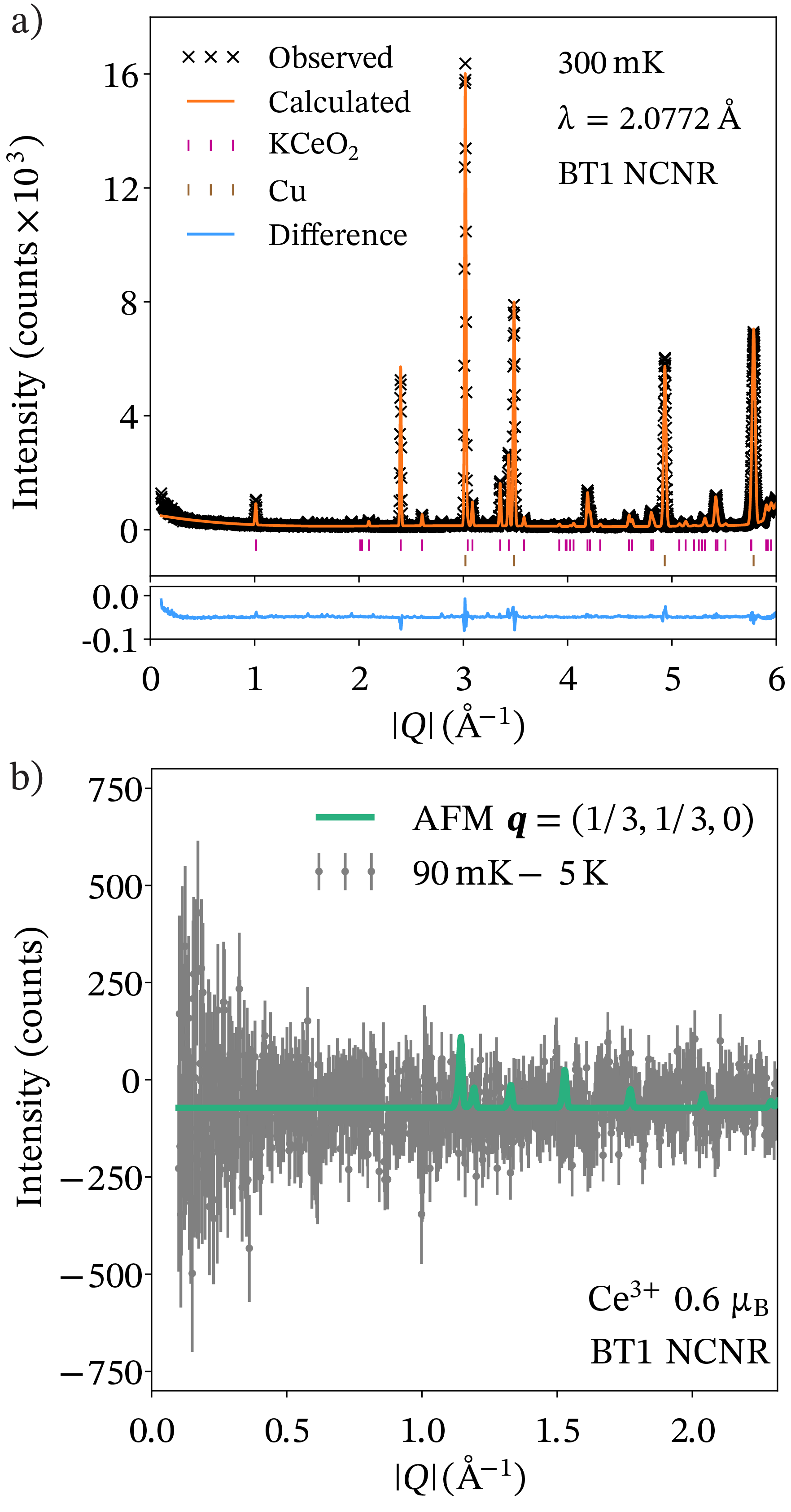}
	\caption{(a) Powder neutron diffraction data of KCeO$_2$ collected at 300 mK on the BT-1 diffractometer. No new magnetic reflections are resolved at 300 mK.  Solid line shows the results of Rietveld refinement of the data and the lower panel shows the difference between the model and the data. (b) 90 mK diffraction data with high temperature (5 K) background subtracted and collected in a dilution refrigerator.  The solid green line shows a simulated diffraction pattern for \textbf{q}=(1/3, 1/3, 0) 120$^{\circ}$ type order with 0.6 $\mu_B$ per Ce site.  It is meant to provide a visual benchmark for the detection limit of magnetic order.  }
	\label{fig:structure_2}
\end{figure}

\section{Experimental Results}

\subsection{Structural and bulk electronic properties}
Neutron diffraction data collected at 300 mK and the refined crystallographic parameters are shown in Figure \ref{fig:structure_1} and Table \ref{tab:tabstruct}. The $R\bar{3}m$ structure of KCeO$_2$ was previously reported at room temperature \cite{clos1970deux}, and this was used as a starting point for the lattice refinement. No chemical site-mixing or vacancies were observed in our analysis to within experimental resolution ($\sim 1$ \%). As illustrated in Fig. 1, there is a sizable trigonal distortion in this material, which will be discussed later when the crystalline electric field level structure is examined.  Furthermore, no magnetic Bragg peaks appear in the refinement at 300 mK, and additional measurements performed at 90 mK similarly failed to resolve magnetic scattering (Fig 2). The difference between 90 mK and 5K diffraction data is shown in Fig. 1 (c).  

\begin{figure}[]
	\includegraphics[scale=.425]{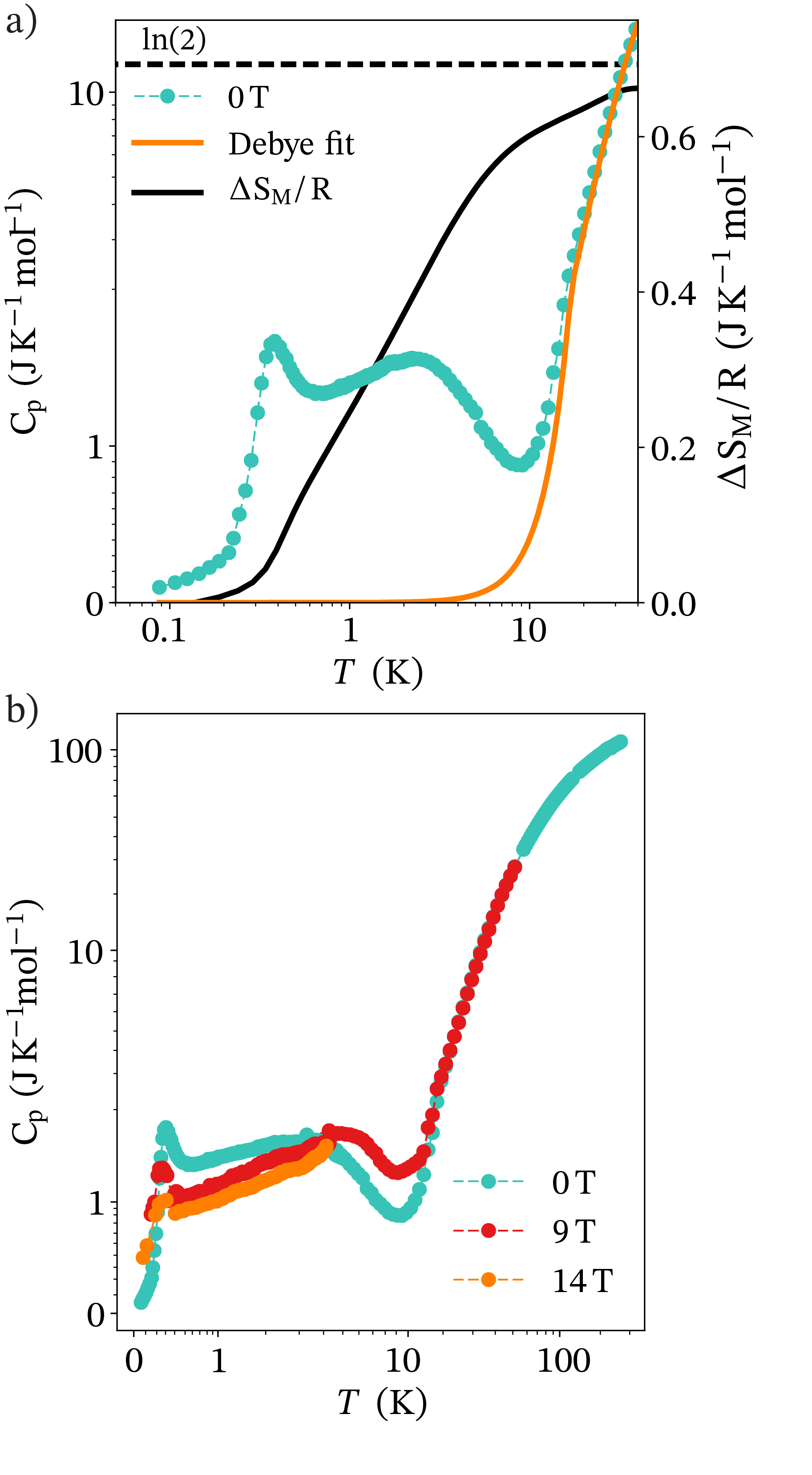}
	\caption{a) Specific heat $C_p(T)$ of KCeO$_2$ (cyan) measured in zero magnetic field and plotted with a Debye fit (orange) modeling lattice contributions. The magnetic specific heat peaks at 300 mK, suggesting KCeO$_2$ enters an ordered state. The integrated magnetic entropy is overplotted as a black line and approaches $R$ln$(2)$ at high temperature (indicated by the horizontal dashed black line). b) $C_p(T)$ of KCeO$_2$ collected under fields of $\mu_0H = 0, 9,$ and $14$ T.}
	\label{fig:cpvfield}
\end{figure}

Specific heat measurements were collected under external magnetic fields of $\mu_0H = 0, 9,$ and $14$ T at temperatures ranging from 80 mK to 300 K. The data are presented in Figure \ref{fig:cpvfield}, and in Figure \ref{fig:cpvfield}a), and two features appear below 10 K. The first is a broad peak centered near $T=3$ K that shifts to higher temperatures with increasing magnetic field strengths and likely indicates the onset of short-range correlations. The second, sharper peak is centered at $T_N =$ 300 mK and is suggestive of the onset of long-range magnetic order. Upon increasing the applied magnetic field, the $T_N =$ 300 mK anomaly softens and shifts to lower temperatures as the Ce moments begin to polarize.

\begin{table}[]
	\caption{Rietveld refinement of structural parameters at 300 mK from elastic neutron scattering data on BT-1. Within error, all ions refine to full occupation and no site mixing is observed. Reduced $\chi^2 =$ 5.23.}
	\begin{tabular}{cc|ccccc}
		\hline
		\multicolumn{2}{c|}{$T$}       & \multicolumn{5}{c}{300 mK}     \\ \hline
		\multicolumn{2}{c|}{$a=b$}     & \multicolumn{5}{c}{3.65861(8) \AA}  \\
		\multicolumn{2}{c|}{$c$}       & \multicolumn{5}{c}{18.58608(85) \AA}  \\ \hline
		Atom           & Wyckoff          & x    & y    & z            & $U_{iso}$ (\AA$^{2}$)      & Occupancy  \\ \hline
		Ce             & 3a            & 0    & 0    & 0            & 0.84(11)   & 0.994(5) \\
		K             & 3b            & 0    & 0    & 0.5          & 1.09(15)   & 0.992(7) \\
		O              & 6c            & 0    & 0    & 0.26939(11)   & 0.97(8)   & 0.999(5) \\ \hline	
	\end{tabular}
	\label{tab:tabstruct}
	
\end{table}

\begin{figure*}[]
	\includegraphics[scale=.375]{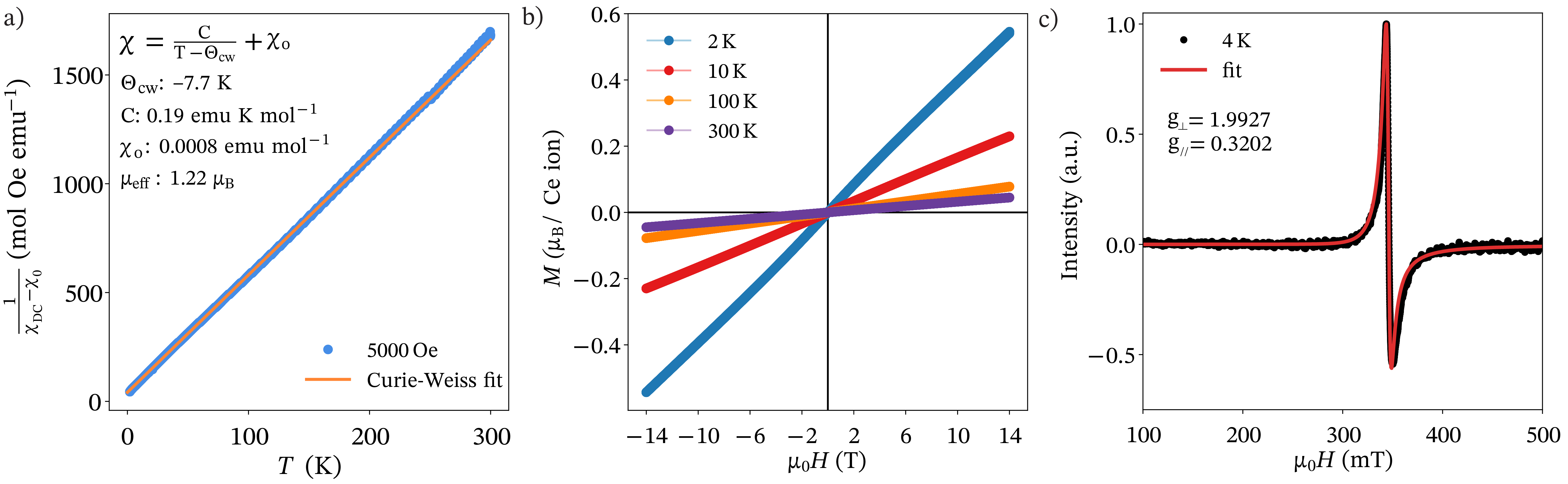}
	\caption{a) Inverse magnetic susceptibility of KCeO$_2$ collected under 0.5 T and overplotted with a Curie-Weiss fit to the data. b) Field dependence of isothermal magnetization at 2, 10, 100, and 300 K. c) EPR data collected at X-band and 4 K fit to a highly anisotropic $g$ factor of $g_{\perp} =$ 1.992 $\pm$ 0.001 and $g_{//} = 0.32$. 1 emu mol$^{-1}$ Oe$^{-1}$ = $4 \pi $10$^{-6}$ m$^{3}$ mol$^{-1}$.}
	\label{fig:suscepmvhesr}
\end{figure*}

To further explore the magnetism at low temperature, magnetic susceptibility and isothermal magnetization measurements were performed.  The resulting data are presented in Figure \ref{fig:suscepmvhesr}. The inverse susceptibility is linear between 2 K to 300 K and was fit to a Curie-Weiss form, yielding antiferromagnetically coupled ($\Theta_{CW} = -7.7$ K) $\mu_{eff} = \sqrt{8C} = 1.22$ $\mu_B$ Ce moments. This corresponds to a powder-averaged $g_{avg}$ factor of $g_{avg} = \mu_{eff}/[\sqrt{J_{eff}(J_{eff}+1)}] = 1.41$ assuming $S_{eff} = 1/2$ Ce moments. Isothermal magnetization data collected at select temperatures are also plotted in Figure \ref{fig:suscepmvhesr}b) and show an approximately linear response from 0 to 14 T. We note that some high-field curvature appears in the 2 K data as the expected saturated magnetization of $\sim 0.7$ $\mu_B$/Ce ion is approached. 

Electron paramagnetic resonance (EPR) data were also collected and are plotted in Figure \ref{fig:suscepmvhesr}c). A highly anisotropic $g$ factor is observed and the signal contains one sharp resonance corresponding to $g_{\perp} = 1.992$ $\pm$ 0.001 and $g_{//} = 0.32$. Broadening of this EPR line shape was attributed to a normal distribution of $g_{//}$ with full-width at half-maximum (FWHM) of 0.12(3). These $g$ factor components indicate $g_{avg} = \sqrt{1/3(g_{//}^2+2g_{\perp}^2)} = 1.63$ $\pm$ 0.01, which is slightly larger than that fit from the inverse susceptibility value. Since EPR is a more direct measurement of the $g$-factor tensor, these EPR values are later used to help model the crystalline electric field level structure.

\subsection{Crystalline electric field excitations}

\begin{figure*}
	\includegraphics[width=\textwidth]{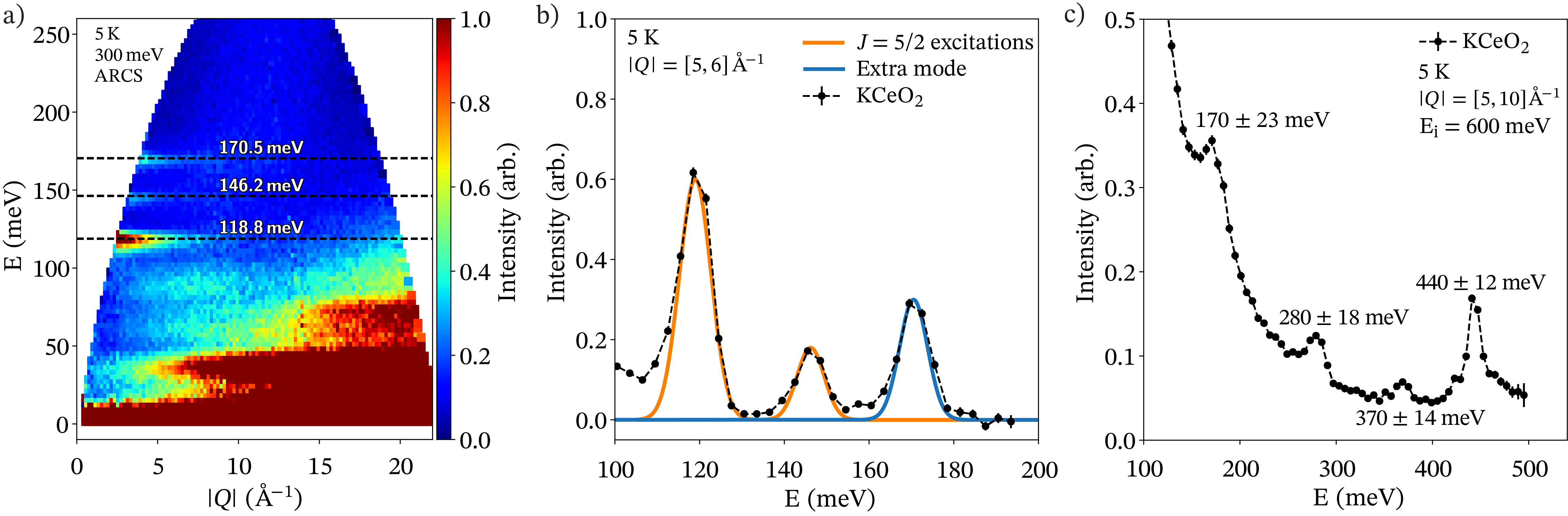}
	\caption{a) Momentum and energy map of neutron scattering intensities collected on the ARCS spectrometer.  b) $\bf{Q}$-averaged energy-cut of inelastic neutron scattering (INS) spectrum $I(\bf{Q},\hbar \omega)$ collected at 5 K and $E_i = 300$ meV on the ARCS spectrometer with full-width-at-half-maximum (FWHM) energy resolution at the elastic line of 12.8 meV. A linear background is subtracted in order to determine integrated intensity ratios. c) $\bf{Q}$-averaged energy-cut of $E_i =$ 600 meV INS data showing the anomalous $E_3$ mode and the onset of transitions into the $J = 7/2$ manifold of states that begins at 280 meV. Error bars represent the FWHM energy-resolution at each energy transfer.}
	\label{fig:CEF}
\end{figure*}

Inelastic neutron scattering data are presented in Figure \ref{fig:CEF} with an incident energy $E_i =$ 300 meV. An uncommonly large splitting of the $J=5/2$ multiplet is observed in Figure \ref{fig:CEF} (a), with the first excited state appearing near 118 meV ---an energy of similar magnitude to the spin orbit coupling strength. The high energy of the crystal field validates our earlier use of the Curie-Weiss model up to 300 K.  Specifically, three sharp excitations with energy widths limited by the resolution of the measurement were observed at $E_1=118.8 \pm 8.0$ meV, $E_2=146.2 \pm 6.9$ meV, and $E_3=170.5 \pm 6.1$ meV where error bars represent the full width at half maximum (FWHM) energy resolution of the instrument (FWHM at $E = 0$ is 12.8 meV).  Data collected with a larger $E_i=600$ meV show excitations into the $J=7/2$ multiplet with transitions at $E_3=280 \pm 18$ meV, $E_4=370 \pm 14$ meV, and $E_5=440 \pm 12$ meV, consistent with the expected spin-orbit coupling strength of Ce$^{3+}$, and revealing three of the four expected $J=5/2$ to $J=7/2$ intermultiplet transitions. To analyze the splitting within the $J=5/2$ multiplet, $Q$-averaged cuts through $I(\bf{Q},\hbar \omega)$ are plotted as a function of energy from the $E_i$ = 300 meV data shown in Figure \ref{fig:CEF} (b). The spectral weights of peaks from this cut were utilized to analyze the CEF level structure.

Notably, within this lower energy data, there is one extra mode beyond those expected for excited doublets within the $J=5/2$ multiplet.  The trivalent Ce$^{3+}$ ions ($4f^1$, $L = 3$, $S = 1/2$) in KCeO$_2$ reside in a local environment with $D_{3d}$ symmetry and total angular momentum $J = |L-S| = 5/2$ according to Hund's rules. In $D_{3d}$ symmetry, the $J = 5/2$ manifold is maximally split into three CEF doublets following Kramers theorem, which should render only two excitations in the INS spectrum. The crystalline electric field level scheme for KCeO$_2$ was recently calculated by \textit{ab initio} modeling by Eldeeb et al. \cite{PhysRevMaterials.4.124001}, and this prediction provides a useful starting point for analyzing the neutron scattering data. To identify the origin of the extra mode, we reference the multireference configuration-interaction with spin-orbit coupling (MRCI+SOC) calculations that predict two $J = 5/2$ intramultiplet excitations near $E=121$ meV and $E=143$ meV, and the first $J = 7/2$ intermultiplet excitation near $E=252$ meV. The INS data are in close correspondence to this level structure, establishing the $E_3 = 170.5$ meV mode as the anomalous outlier. Therefore, for the purposes of analyzing the CEF ground state, the $E_3$ mode was not utilized to fit the CEF Hamiltonian.  The origin of this mode is currently unknown and is discussed in further detail in later paragraphs. 

Using the parameterized $E_1$ and $E_2$ excitations, the neutron data can be fit to model the CEF Hamiltonian.  A $D_{3d}$ Hamiltonian of CEF parameters $B_n^m$ and Steven's operators $\hat{O}_m^n$ \cite{StevensOperators} becomes:

\begin{equation}
	\label{eq:CEF}
	H_{CEF} = B_2^0 \hat{O}_2^0 + B_4^0 \hat{O}_4^0 + B_4^3 \hat{O}_4^3
\end{equation}

The CEF Hamiltonian was diagonalized in the CEF interface of Mantid \cite{Mantid} to determine energy eigenvalues and eigenvectors of the $J = 5/2$ excited states. Intramultiplet transition probabilities and $g$ tensor components were calculated from the resulting wave functions. These values were then fit to integrated intensity ratios of the two excitations in the INS data and the $g$-factor components from EPR data by following the minimization procedure reported in Bordelon et al. \cite{PhysRevB.103.014420}. The results are presented in Table \ref{tab:tabCEF}, and the final fit of the CEF scheme contains a superposition of $m_j = 1/2$ and $m_j = 5/2$ angular momentum states that yield anisotropic $g$ factor components of $g_{//} = 0.2910$ and $g_{\perp} = 1.9973$. The CEF parameters obtained from the fit are $B_2^0 = 3.7740$, $B_4^0 = -0.2120$, and $B_4^3 = 6.3963$.

The best fit to the data significantly overestimates the intensity ratio $I_2/I_1$, which likely indicates some of the expected spectral weight is split and appears as part of the extra $E_3$ meV mode. Adding the normalized areas of the $E_2$  and $E_3$ modes ($I_2/I_1 + I_{e}/I_1 = 0.699$) gives a value closer to the value predicted in the best CEF model in Table \ref{tab:tabCEF} (resulting in an improved $\chi^2$=0.0097).  It is worth commenting on the potential origins of the additional 170.5 meV $E_3$ mode.  Spin-orbit coupling is expected to split the $J = 5/2$ multiplet from the $J = 7/2$ manifold by at least 250 meV \cite{PhysRevMaterials.4.124001}, consistent with our observation of an intermultiplet excitation at $E=252$ meV.  This considerably weakens the possibility that the extra mode at 170 meV arises from a transition into a $J=7/2$ state. Anomalous CEF modes observed for lanthanide ions in similar crystallographic environments are often explained by invoking coupling to phonons or vibronic effects \cite{thalmeier1982bound, thalmeier1984theory, vcermak2019magnetoelastic, PhysRevLett.108.216402, PhysRevB.92.144422, PhysRevB.55.180, PhysRevB.60.R12549}, conjectures of hydrogen embedded within the sample \cite{rush1980neutron, PhysRevB.55.5700, PhysRevLett.122.187201, gao2019experimental, sibille2020quantum, PhysRevLett.115.097202, bastien2020long}, or the potential presence of chemical impurities creating multiple local environments \cite{PhysRevLett.108.216402, PhysRevB.92.144422, PhysRevB.55.180, PhysRevB.60.R12549, bastien2020long, li_zhang4, PhysRevB.92.134420, PhysRevB.97.024415}. Somewhat uniquely, none of these explanations adequately explain the origin of the $E_3$ mode in KCeO$_2$. 

First, while the Ce wave function is known to strongly couple to the lattice, typical vibronic coupling occurs when a CEF mode resides close in energy to a phonon branch. In this scenario, the phonon and CEF mode couple together to create a bound state and a splitting or broadening of a CEF mode results. This has been observed in various lanthanide materials (e.g. CeAl$_2$ \cite{thalmeier1982bound, thalmeier1984theory}, CeCuAl$_3$ \cite{vcermak2019magnetoelastic, PhysRevLett.108.216402}, LiYbF$_4$ \cite{PhysRevB.92.144422}, and YbPO$_4$ \cite{PhysRevB.60.R12549}); however, the intramultiplet splitting in KCeO$_2$ is unusually high in energy, well above the single phonon cutoff apparent in our INS measurements. Additionally, phonon-induced splitting of a CEF mode typically creates two excitations above and below the unhybridized CEF energy. INS data in KCeO$_2$ resolve two CEF modes at $E_1 = 118.8$ meV and $E_2 = 146.2$ meV, each remarkably close to the calculated energies from Eldeeb et al. \cite{PhysRevMaterials.4.124001}, strongly suggesting the $E_3 = 170.5$ meV mode is not a result of trivial splitting of either $E_1$ or $E_2$ excitations.

\begin{table}[]
	
	\caption{The fit CEF wave functions for KCeO$_2$ determined from minimizing parameters extracted from $E_i$ = 300 meV INS data and EPR $g$ factor components. The energy level scheme from Eldeeb et al. \cite{PhysRevMaterials.4.124001} calculated with multireference configuration-interaction and spin-orbit coupling (MRCI+SOC) was used as a starting point and the 170.5 meV excitation was excluded from this analysis.}
	\begin{tabular*}{9.0cm}{l|llllll|l}
		\hline
		& $E_1$   & $E_2$    & $\frac{I_2}{I_1}$   & $g_{avg}$ & $g_{\perp}$ & $g_{//}$  & $\chi^2$   \\ \hline	
		Fit  & 118.8 & 145.7 & 0.639 & 1.6395 & 1.9973 & 0.2910 & 0.570         \\ \hline 
		Observed   & 118.8 & 146.2  & 0.257 & 1.6340 & 1.992 & 0.32 &     
	\end{tabular*}
	
	
	\hskip+0.1cm
	\begin{tabular*}{9.025cm}{lll}
		\hline
		Fit wave functions: & & \\
		$| \omega_{0, \pm}\rangle =$  $0.881|\pm1/2\rangle + 0.473|\mp5/2\rangle$ & &\\		
		$| \omega_{1, \pm}\rangle =$  $1|\pm3/2\rangle$ & &\\
		$| \omega_{3, \pm}\rangle =$  $ - 0.473|\mp1/2\rangle  \mp 0.881|\pm5/2\rangle$ & &\\ \hline
	\end{tabular*}
	\label{tab:tabCEF}
	
\end{table}

\begin{figure*}[]
	\includegraphics[width=\textwidth]{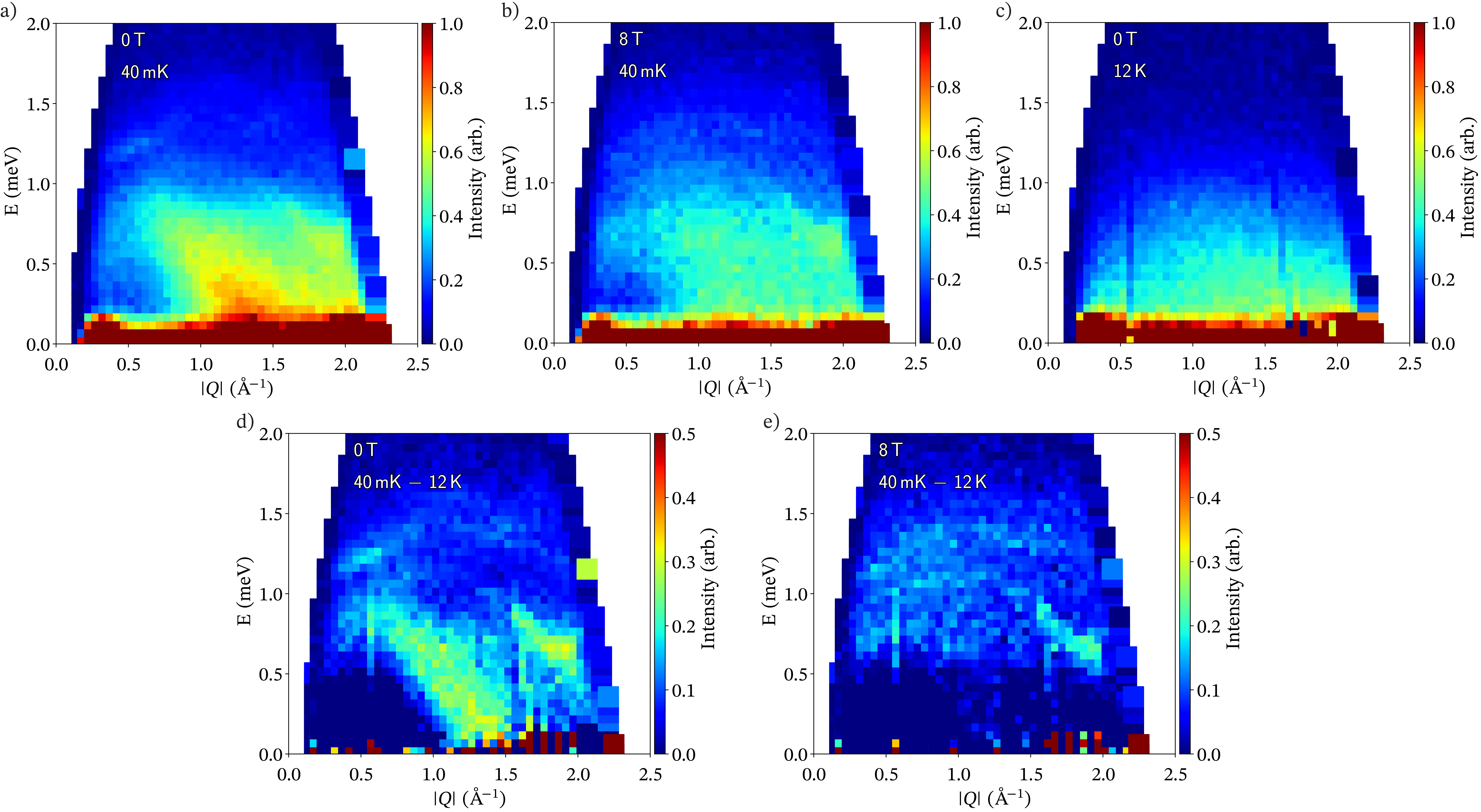}
	\caption{Low energy inelastic neutron scattering (INS) spectra $I(\bf{Q},\hbar \omega)$ collected on the CNCS spectrometer at a) $\mu_0H =$ 0 T and 40 mK, b) $\mu_0H =$ 8 T and 40 mK, c) $\mu_0H =$ 0 T and 12 K. Magnetic intensity suggestive of static order originates from $\bf{Q} =$ 1.4 \AA, close to the crystallographic $K$ point. d) Subtracting $\mu_0H =$ 0 T and 12 K data from the data in panel a) reveals spin wave-like excitations in KCeO$_2$. Two distinct modes appear, the first with a bandwidth up to 1 meV from the structural $K$ point and the second, flatter mode with its spectral weight anchored close to 1.25 meV. e) Under a field of $\mu_0H =$ 8 T, the powder-averaged modes begin to diffuse.}
	\label{fig:lowINS}
\end{figure*}

Second, while hydrogen impregnation is another potential source of unexplained CEF transitions in materials \cite{rush1980neutron, PhysRevB.55.5700, PhysRevLett.122.187201, gao2019experimental, sibille2020quantum, PhysRevLett.115.097202, bastien2020long}, the synthesis of KCeO$_2$ requires strict atmospheric handling in an inert glove box environment.  A \textit{significant} concentration of hydrogen would be required to generate the spectral weight at $E_3$ and is unlikely to generate the sharp energy line width observed. Related to this notion, undetected chemical impurities would arise in other experimental artifacts other than one extra INS excitation. They typically create multiple chemical environments surrounding Ce ions and generate shifted $J = 5/2$ intramultiplet excitation schemes that broaden CEF excitations or simply alter the Ce ion valence. Neither of these scenarios result in only one additional sharp CEF mode. Even a separate $J = 5/2$ Ce$^{3+}$ environment with perfect cubic symmetry at the oxygen coordination level is expected to have a split quartet due to the farther neighbor anisotropic environment \cite{PhysRevMaterials.4.124001}. In all cases, the concentration of any secondary environment would be well within the elastic neutron scattering refinement resolution in order to create the intensity observed within the $E_3$ mode. In other words, the elastic neutron scattering resolution rules out chemical impurities to within 1\%, and a 1\% impurity could not account for the entire spectral intensity at $E_e =$ 170.5 meV in INS data.

To further demonstrate this, the most likely defect state---Ce ions within the K planes---was considered via an embedded-cluster material model and quantum chemical CASSCF+SOC computations.  The calculations show excited state energies at 78.5 meV and 83.5 meV due to the ligand field about the defect state.  These energies are far below that of the $E_3$ mode, and trivial point-charge models incorporating vacancies similarly show excitation energies far below those observed experimentally.  This combined with the constraints imposed by diffraction data strongly argue against localized chemical/lattice impurities as the origin of the $E_3$ mode.

\subsection{Low-energy magnetic excitations}

Low-energy inelastic neutron scattering data are plotted in Figure \ref{fig:lowINS}, which displays both the field and temperature dependence of low-energy spin dynamics collected at 40 mK, 12 K, $\mu_0H =$ 0 T, and $\mu_0H =$ 8 T. Despite no resolvable magnetic Bragg peaks appearing in the elastic neutron scattering data, magnetic spectral weight appears at finite frequencies in the zero field 40 mK data near $|Q| = 1.4$ \AA \, close to the magnetic zone centers (1/3, 1/3, $L=0-4$) for three sublattice order and with a bandwidth of roughly 1.5 meV. Using the zero field, 12 K data as an approximate paramagnetic background, subtracted data are plotted in Figure \ref{fig:lowINS}d) and reveal signatures of two spin wave branches. The first contains gapless modes dispersing up to $E=1$ meV originating near $|Q| = 1.4$ \AA, and the second appears as a gapped branch centered near 1.25 meV with a smaller bandwidth. Under an external field of $\mu_0H = 8$ T, the low energy fluctuations at 40 mK are suppressed as Ce moments begin to polarize, and the low energy gapless modes are quenched.   We note here that, due to the difficulty in thermalizing powder samples, 40 mK is only the nominal temperature of the KCeO$_2$ sample in this experiment. We expect that the actual sample temperature is higher, with 100 mK being a reasonable estimate based on prior experience.  

\section{Discussion}

Our aggregate data suggest the formation of a magnetically ordered state below 300 mK in KCeO$_2$.  While an antiferromagnetic $\Theta_{CW} = -7.7$ K from Curie-Weiss analysis indicates a strong degree of magnetic frustration, a second, sharp peak in the low-temperature specific heat combined with low-energy INS data showing multi-branch spin excitations at 40 mK indicate a magnetically ordered ground state. This observation is consistent with a recent report of long-range magnetic order in the sulfur-based analog KCeS$_2$ \cite{bastien2020long}. Although no new magnetic Bragg peaks were observed in elastic neutron scattering data collected at 300 mK for KCeO$_2$, the likely reason is that the ordered moment is below the limit detection in current measurements.  

One naively expects that the saturated ordered moment is near the limit of detection.  NaCeO$_2$, for instance, develops a $\mu=0.57$ $\mu_B$ ordered antiferromagnetic structure below 4 K \cite{PhysRevB.103.024430}, suggesting an ordered moment in KCeO$_2$ of similar magnitude. Simulation of a test 120$^{\circ}$, ${\bf{k}} = (1/3, 1/3, 0)$ antiferromagnetic structure with an ordered moment $\mu=0.6$ $\mu_B$ shown in Fig. 2 reveals that any magnetic Bragg scattering from this structure is outside of the current experiment's resolution and signal to noise ratio. As an upper bound, the powder averaged $g_{avg}$=1.6375 from EPR measurements implies that the maximal ordered moment in KCeO$_2$ is only $g_{avg} J_{eff}\mu_B = 0.82$ $\mu_B$. A moderate reduction from this maximal value due to quantum fluctuations or perhaps distributed intensity via short-range correlations could easily push powder-averaged magnetic scattering below the limits of detection in our current experiments.

The low-energy INS data on the other hand reveal spin wave-like modes indicative of an ordered ground state. Unfortunately, the coarser momentum resolution of the INS measurement combined with increased background from the magnet sample environment precludes resolution of any magnetic Bragg peaks in the elastic channel of the data. The lower branch of the spin waves originate close to the $K$ point of KCeO$_2$ expected for ${\bf{k}} = (1/3, 1/3, 0)$-type antiferromagnetic order. Though, absent a direct observation of magnetic Bragg scattering, we are unable to further constrain the potential ordered state. 

It is interesting to contrast the magnetic ground states of triangular lattice $ARX_2$ ($A =$ alkali, $R =$ rare earth, $X =$ chalcogenide) systems with $S_{eff}$=1/2 moments built from a single hole in the $4f$ shell ($R$ = Yb$^{3+}$) and for those built from a single electron ($R$ = Ce$^{3+}$).  While NaYbO$_2$ shows no signs of long-range magnetic order down to 50 mK and a continuum of spin excitations resembling a quantum spin liquid \cite{bordelon2019field, PhysRevB.101.224427, baenitz_doert, ding_tsirlin, ranjith_baenitz}, KCeO$_2$ seemingly shows signs of ordered magnetism forming below 300 mK. Although the two compounds share an identical lattice framework of frustrated $S_{eff} = 1/2$ moments on a delafossite-based triangular lattice, the differing anisotropies of Ce$^{3+}$ and Yb$^{3+}$ may account for the differences in low temperature ordering. As one example, Ce$^{3+}$ moments in KCeO$_2$ exhibit strongly anisotropic $g$-tensors, while Yb$^{3+}$ moments have a reduced anisotropy of $g_{//} = 1.73$ and $g_{\perp} = 3.29$ \cite{bordelon2019field}. Important differences may also exist in relation to the intersite exchange anisotropies, as a result of different dominant superexchange processes in the case of $f^1\!-\!p^6\!-\!f^1$ and $f^{13}\!-\!p^6\!-\!f^{13}$
electron configurations.  Understanding why varying the trivalent lanthanide in this $R\bar{3}m$ delafossite structure induces ordering in Ce-based variants and a quantum disordered state in Yb-based variants will require deeper investigation into the ordered state in KCeO$_2$ and related compounds such as KCeS$_2$ \cite{bastien2020long}. 

Trivalent Ce$^{3+}$ ions in a $D_{3d}$ environment are proposed to contain a dipole-octupole Kramers doublet ground state \cite{GChen1, GChen2, gao2019experimental, PhysRevLett.122.187201, sibille2020quantum, PhysRevLett.115.097202} of pure $m_j = 3/2$ character. Separate dipolar and octupolar order can occur from this unique wave function and has been proposed for both the triangular lattice \cite{GChen1} and pyrochlore lattice \cite{GChen2}. While this ground state has been observed in pyrochlores Ce$_2$Sn$_2$O$_7$ \cite{sibille2020quantum, PhysRevLett.115.097202} and Ce$_2$Zr$_2$O$_7$ \cite{gao2019experimental, PhysRevLett.122.187201}, it has yet to be reported on the triangular lattice where a rich phase space of states is predicted for both the octupolar and dipolar components \cite{GChen1}. 

In KCeO$_2$, such a dipole-octopole doublet is not favored and our crystal field analysis shows that the ground-state wave function is far from a pure $m_j = 3/2$ Kramers doublet. Table \ref{tab:tabCEF} shows the CEF wave functions with the $m_j = 3/2$ Kramers doublet comprising the first excited state at $E_1 = 118.8$ meV. Instead, the ground state wave function is a normal Kramers dipole doublet with mixed $m_j = 1/2$ and $m_j = 5/2$ character, and the symmetry of the wave functions determined from our CEF analysis agrees with predictions from Eldeeb \textit{et al.} \cite{PhysRevMaterials.4.124001}. In comparison, the pyrochlores Ce$_2$Sn$_2$O$_7$\cite{sibille2020quantum, PhysRevLett.115.097202} and Ce$_2$Zr$_2$O$_7$\cite{gao2019experimental, PhysRevLett.122.187201} contain two extra O$^{2-}$ anions surrounding each Ce$^{3+}$ ion, which seemingly favor a dipole-octupole Kramers doublet ground state \cite{PhysRevMaterials.4.124001}. 

The excitations in KCeO$_2$, however, do reveal an anomalous mode intermediate to the end of the predicted intramultiplet $J=5/2$ excitations and the onset of the first $J=5/2$ to $J=7/2$ transition. This extra Ce$^{3+}$ $D_{3d}$ mode is not limited to KCeO$_2$. A similar type of high energy extra CEF mode is also present in other reported materials with similar $D_{3d}$ environments, KCeS$_2$ \cite{bastien2020long} and Ce$_2$Zr$_2$O$_7$ \cite{gao2019experimental, PhysRevLett.122.187201}. In these materials, the extra mode was relatively weak in comparison to the other two excitations, and, as a result, the extra modes were accounted for via suppositions of weak chemical impurities or hydrogen incorporation. Additionally, Ce$_2$Sn$_2$O$_7$ \cite{sibille2020quantum, PhysRevLett.115.097202} reported weak extra CEF modes within the $J = 5/2$ manifold that may also be related to the origin of the single extra mode of KCeO$_2$. For KCeO$_2$, the third unaccounted excitation at $E_3 =$ 170.5 meV is intense and energetically well-separated from the other two excitations $E_1$ and $E_2$. Its origin remains unknown and may be linked to the extra INS modes in KCeS$_2$, Ce$_2$Zr$_2$O$_7$, and Ce$_2$Sn$_2$O$_7$, who share a common Ce$^{3+}$ in a $D_{3d}$ CEF environment. 

One potential origin of the $E_3$ mode is the presence of a strong dynamic Jahn-Teller effect akin to that observed in isovalent PrO$_2$ \cite{PhysRevLett.86.2082}.  Coupling to vibronic modes associated with this effect can reduce the ordered moment and redistribute magnetic intensity into localized modes not captured within static models of the CEF spectrum.  The $E_3$ mode in KCeO$_2$, however, is reasonably sharp in energy (resolution-limited), and there is no broad distribution of magnetic spectral weight elsewhere expected from coupling to an array of lattice modes.  This places rather stringent constraints on a potential vibronic mechanism and merits further systematic study.

\section{Conclusions}

We explored the magnetic ground state of KCeO$_2$, which crystallizes in an ideal $R\bar{3}m$ triangular lattice antiferromagnetic structure with $S_{eff} = 1/2$ Ce$^{3+}$ ions. The material develops signatures of magnetic ordering below $T_N = $ 300 mK. Below $T_N$, spin wave-like excitations of bandwidth $\approx$ 1.5 meV appear in low-energy inelastic neutron scattering data and originate near $|Q| = 1.4$ \AA$^{-1}$. High-energy inelastic neutron scattering data reveal a strong intramultiplet crystalline electric field splitting of the Ce$^{3+}$ $J = 5/2$ states with levels identified closely matching previous calculations \cite{PhysRevMaterials.4.124001}. However, an unexplained extra local, orbital excitation appears that cannot be accounted for via conventional impurity/phonon mechanisms and whose presence should be further investigated in related materials. Understanding the origin of this unconventional mode and the microscopic distinction between the ordered state in KCeO$_2$ and the quantum disordered state in NaYbO$_2$ are important questions motivating the continued exploration of $ARX_2$ ($A =$ alkali, $R =$ rare earth, $X =$ chalcogenide) $R\bar{3}m$ materials.

\begin{acknowledgments}
	S.D.W. and M.B. sincerely thank Leon Balents and Chunxiao Liu for helpful discussions.  This work was supported by the US Department of Energy, Office of Basic Energy Sciences, Division of Materials Sciences and Engineering under award DE-SC0017752 (S.D.W. and M.B.).  Research reported here also made use of shared facilities of the UCSB MRSEC (NSF DMR-1720256). A portion of this research used resources at the Spallation Neutron Source, a DOE Office of Science User Facility operated by Oak Ridge National Laboratory.  The work at Georgia Tech (heat-capacity measurements) was supported by the NSF under NSF-DMR-1750186.  M.S. and X.W. acknowledge support via the UC Santa Barbara NSF Quantum Foundry funded via the Q-AMASE-i program under award DMR-1906325.  T.P. and L.H. acknowledge financial support from the German Research Foundation (grants HO-4427/3 and HO-4427/6).
	M.S.E, T.P., L.H., and U.K.R. thank Ulrike Nitzsche for technical assistance.  Certain commercial equipment, instruments, or materials are identified in this document. Such identification does not imply recommendation or endorsement by the National Institute of Standards and Technology, nor does it imply that the products identified are necessarily the best available for the purpose.
\end{acknowledgments}

\bibliography{KCO_bib_v1}

\end{document}